\documentclass[aps,showpacs,prd,amsmath,twocolumn,amssymb,floatfix,superscriptaddress]{revtex4}

\usepackage{epsfig}
\usepackage{graphicx,psfrag}
\usepackage{dcolumn}% Align table columns on decimal point
\usepackage{bm}

\newcommand{\sla}[1]{\mbox{$#1\!\!\!/$}}
\newcommand{\beq}{\begin{equation}}
\newcommand{\eeq}{\end{equation}}
\newcommand{\bea}{\begin{eqnarray}}
\newcommand{\eea}{\end{eqnarray}}
\newcommand{\hf} {\frac{1}{2}}
\newcommand{\nonu}{\nonumber\\}
\newcommand{\nn}{\nonumber\\}

\newcommand\eqn[1]     {Eq.\,(\ref{#1})}
\newcommand\eqns[2]    {Eqs.\,(\ref{#1}) and~(\ref{#2})}

\newcommand\fig[1]     {Fig.\,{\ref{#1}}}
\newcommand\sect[1]    {Sect.\,{\ref{#1}}}

\def\tu{{\tilde u}}

\def\ord#1{{\cal O}(#1)}

\def\mr#1{{\mathrm{#1}}}
\def\ci{{\rm i}}

\begin{document}

\title{Generalized universality in the massive sine-Gordon model}

\author{S. Nagy}

\affiliation{Department of Theoretical Physics, University of Debrecen,
Debrecen, Hungary}

\author{I. N\'andori}

\affiliation{Institute of Nuclear Research of the Hungarian Academy of Sciences,
H-4001 Debrecen, P.O.Box 51, Hungary}

\author{J. Polonyi}

\affiliation{Institute for Theoretical Physics, Louis Pasteur University,
Strasbourg, France}

\author{ K. Sailer}

\affiliation{Department of Theoretical Physics, University of Debrecen,
Debrecen, Hungary}

\date{\today}

\begin{abstract}
A non-trivial interplay of the UV and IR scaling laws, a generalization of the
universality is demonstrated in the framework of the massive sine-Gordon model,
as a result of a detailed study of the global behaviour of the renormalization
group flow and the phase structure.
\end{abstract}


\maketitle

\section{Introduction}
Our hope to cover the overpowering richness of Physics by microscopic theories
constructed on simple elementary principles is based on the concept of 
universality, the possibility of ignoring most of the short distance, 
microscopic parameters of the field theoretical models in describing the
dynamics at finite scales \cite{univ}. Such an enormous simplification,
obtained by inspecting the asymptotical UV scaling laws, is possible if the
renormalized trajectory `spends' several orders of magnitude in the UV scaling
regime and suppresses the sensitivity of the physics on the
non-renormalizable, i.e. irrelevant parameters.

But one tacitly assumes in this scenario that there are no other scaling
laws in the theory or at least they do not influence the results obtained
in the UV scaling regime. This is certainly a valid assumption for models
with explicit mass gap in the Lagrangian which renders all non-Gaussian
operator irrelevant at the IR fixed point. But what happens in theories
with massless bare particles? Spontaneous symmetry breaking, or
dynamical mass generations in general, may change the situation
\cite{grg,janosRG} and open the possibility of tunable parameters of the
theories which make their impact on the dynamics due to a competition between
the UV and the IR scaling regimes. Such a competition can be highly
non-trivial because both of these regimes can cover unlimited orders of
magnitude of evolution for the RG flow and may offer a new way of
looking at complex systems.

The traditional way of 'understanding' a
large system starts with the identification of its components and their 
elementary behaviour and then continues with the combination of these rules
to come up with the whole picture. The competition between an UV and IR fixed
points may introduce microscopical parameters which
influence the dynamics in the macroscopic range only. Such a 
renormalization microscope mechanism allows us to determine certain microscopical 
parameters by means of the long range dynamics and represents a microscopic,
elementary feature of a model which remains undetected unless the whole 
system is considered.

We turn to a simple two-dimensional model in this work whose dynamics displays
such a phenomenon. It has an explicit mass gap but this does not render the IR
dynamics  uninterestingly suppressed. This is because the natural strength of
a coupling constant is its value when expressed in units of the scale inherent
of the phenomenon to be described, e.g. the gliding cutoff in the framework of the
renormalization group (RG) in momentum space. Since  the non-Gaussian coupling
 constants of the
local potential have positive mass dimension in two space-time dimensions, the
usual slowing down of the evolution of dimensional coupling constants below
the mass gap generates diverging dimensionless coupling constants, i.e.
relevant parameters at the IR fixed point. But this is one part of the story
only. The other part is that in order to have non-trivial competition between
the UV and IR scaling regimes we need coupling constants which are
non-renormalizable, i.e. irrelevant at the UV fixed point and turn into
relevant at the IR end point. In order to render the coupling constants of the
local potential non-renormalizable we need non-polynomial coupling. The
simplest, treatable case is the sine-Gordon (SG) model \cite{Coleman_sg} whose Lagrangian
contains a periodic local potential and which possesses two phases. All
coupling constants of the local periodic potential are non-renormalizable in
one of the phases when the model is considered in the continuous space-time
with non-periodic space-time derivatives \cite{huang}. 
We are thus lead to the massive sine-Gordon (MSG) model
\cite{Coleman,Ichinose,Lu2000,nagys_jpa,nandori_npb} which has already been
thoroughly investigated in the seventies as the bosonized version of the
massive Schwinger model, 1+1 dimensional QED, the simplest model possessing
confining vacuum \cite{Coleman}. The different UV and IR scaling laws
have already been addressed in this model \cite{Ichinose,nagys_jpa} but the
careful identification of the full scaling regime, not only the asymptotical
ends is needed to understand the global features of the RG flow diagram. This
is the goal of the present work. One can only answer  this question
by solving the RG equations for the couplings and to find the effective
potential. In this work we use the Wegner-Houghton (WH) functional RG method 
\cite{wh},
where the renormalization procedure is defined by the blocked action with
gliding sharp cut-off to determine the flow of the Fourier amplitudes of the model.
The handicap of the WH-RG method is that one cannot go beyond (if it even necessary)
the local potential approximation (LPA), since
the gradient expansion of the Lagrangian gives indeterminate
evolution equation due to the sharp cut-off used. There are
several other methods in the literature which are based on the
evolution of the effective action \cite{effac_rg}. We show in this paper that the
internal space RG method \cite{internal}, where the RG evolution is controlled by the mass and
allows us in principle to go beyond the LPA gives the same results in LPA
as the WH-RG method.

Since the SG model
has a condensed phase which arises at a finite cut-off $k_\mr{SI}$
one expects that if $M<k_\mr{SI}$ the condensation also should appear
in the MSG model. In our previous works \cite{nagys_jpa,nagys_plb} we showed in the
WH-RG framework that the
non-trivial scalings appear in the deep IR limit for the SG and the MSG models, indeed.
Since the SG model has a trivial, constant effective potential in either phase
\cite{nandori,nagys_plb}, 
the RG methods based on the effective action \cite{internal,effac_rg} may fail in
treating both the SG and the MSG models. Nevertheless,
when the control parameter approaches the  to  physical value of the mass,
the internal-space RG  enables one to find
an evolution of the parameters which is analogous to their WH-RG flow. 
Also the sign of spinodal instability seems to appear as a singularity in the internal-space 
evolution. In the latter case the analogy mentioned above allows one to change from the 
internal-space RG analysis to the WH-RG framework and to determine the phase structure
of the MSG model.  The situation seems to be similar as that for the WH-RG framework where
the IR limit $k\to0$  of the blocked action is trivial but physics can be read off
from its approaching this limit.

A side-product of our RG analysis is that one recovers the well-known phase structure of the
bosonized version of QED$_2$
in the LPA. Therefore, one is lead to the conclusion that the phase structure, as well as the
spinodal instability should survive wave-function renormalization.

Another possibility might be the usage of the Polchinski RG method
which uses the blocked action and a smooth cut-off. Unlike the WH-RG
method  based on summing up the loop corrections during the
evolution, Polchinski's procedure sums up the perturbation
expansion, therefore it misses the spinodal instability
in an obvious manner.
In fact, one can easily show that the local potential evolved by this scheme is not of
parabolic shape, a feature 
is thought to be a crucial sign of spinodal instabilities in LPA.

The ionized (large $\beta^2$) phase of the MSG model
can be an example where thorough numerical investigations show
that the bare UV irrelevant coupling becomes relevant in the deep IR regime.
This fact clearly shows that the treatments of the MSG model
based on any perturbation expansion is doomed to failure.

The paper is organized as follows. In \sect{sec:blocking}
we derive the evolution of the coupling constants of the MSG model in the framework of the 
WH-RG method. In \sect{sec:phase} the phase structure of the MSG model is presented.
The RG microscope effect is discussed in \sect{sec:mic}. The evolution of the couplings
constants in the internal-space RG framework is treated in \sect{sec:internal}.
We show in \sect{sec:qed} that our WH-RG results recover the well-known phase
structure of QED$_2$, and finally, in \sect{sec:sum} the conclusion is drawn up.

\section{Blocking in momentum space}
\label{sec:blocking}

The MSG model is defined in 2-dimensional, infinite, Euclidean space-time by
the Lagrangian
\beq
S_k = \int_x\left[\frac12(\partial_\mu\phi_x)^2+U_k(\phi_x)\right],
\eeq
given in the leading order, local potential approximation (LPA) of the
gradient expansion, $k$ denotes the sharp momentum space cutoff and the
potential is the sum $U_k(\phi)=\hf M^2_k\phi^2+V_k(\phi)$, the second term
being periodic,
\beq
V_k (\phi)= \sum_{n=1}^\infty u_n(k)\cos(n\beta_k\phi).
\label{per}
\eeq
The blocking in momentum space \cite{nagys_plb}, the lowering of the cutoff, $k\to k-\Delta k$, 
consists of the splitting the field variable, $\phi=\tilde\phi+\phi'$ in such
a manner that $\tilde\phi$ and $\phi'$ contains Fourier modes with
$|p|<k-\Delta k$ and $k-\Delta k<|p|<k$, respectively and the integration
over $\phi'$ leads to the WH equation \cite{wh}
\beq
\label{WHdim}
\left(2 + k\partial_k \right) {\tilde U}_k ( \phi) = 
-\frac1{4\pi}\ln\left(1 + {\tilde U}''_k (\phi) \right)
\eeq
for the dimensionless local potential ${\tilde U}_k = k^{-2} U_k$. This
equation is obtained by assuming the absence of instabilities for the modes
around the cutoff. Only the WH-RG scheme which uses a sharp gliding  cutoff
can account for the spinodal instability, which appears when the restoring
force acting on the field fluctuations  to be eliminated is vanishing and
the resulting condensate generates tree-level contributions to the evolution
equation \cite{janostree}. The saddle point for the blocking step, $\phi'_0$, 
is obtained by minimizing the action, 
$S_{k-\Delta k} [\phi] = \min_{\phi'} \left(S_k[\phi + \phi'] \right)$
\cite{janosRG,nandori}. The restriction of the minimisation for plane waves
gives
\beq
\label{treedim}
{\tilde U}_{k-\Delta k}(\phi) = \min_\rho \left[\rho^2 +\hf
\int_{-1}^1 du {\tilde U}_k(\phi + 2\rho \cos(\pi u)) \right]
\eeq
in LPA where the minimum is sought for the amplitude $\rho$ only.

One can show that both evolution equations, \eqns{WHdim}{treedim}, preserve
the period length of the potential $V_k(\phi)$ and the non-periodic part 
of the potential, therefore $M_k^2=M^2$ and $\beta_k=\beta$. 
Thus the mass is relevant parameter of the LPA ansatz for all scales,
\beq
\tilde M^2_k=\tilde M^2_\Lambda\left(\frac{k}\Lambda\right)^{-2}.
\label{dlmassev}
\eeq

\subsection{Asymptotic scaling}
It is easy to find the asymptotic UV and IR scaling laws. One finds the
evolution equation
\beq
k\partial_k\tilde u_n(k)=\left(\frac{\beta^2n^2}{4\pi(1+\frac{M^2}{k^2})}
-2\right)\tilde u_n(k)
\label{mcoruv}
\eeq
in the UV regime after ignoring $\ord{M^2/k^2}$ and
$\ord{|U''_\Lambda|^2/k^2}$ contributions with the solution
\beq
\label{nonlinsol1}
\tu_n(k) = \tu_n(\Lambda)\left(\frac{k}{\Lambda}\right)^{-2}
\left(\frac{k^2 + M^2}{\Lambda^2 + M^2}\right)^{\beta^2 n^2/8\pi}.
\eeq
The asymptotic IR scaling, well below the mass scale, is trivial because the
mass gap freezes all scale dependence. The numerical solution of the complete
evolution equation \eqn{WHdim} is shown in \fig{fig:sgmsg}.

\subsection{Impact of the mass gap}

It is instructive to compare the RG flow of the (massless) SG and the MSG
models. The asymptotic UV evolution equations differ in $\ord{M^2/k^2}$ terms
only and the mass term gives small corrections to the scaling laws in this
regime. But the mass gap freezes out the evolution for any values of $\beta$
thus more significant differences should show up between the SG and the MSG
models. The dimensionful potential approaches a constant in the SG model as a
result of the loop-generated or instability driven evolution in the ionized
or the molecular phase, respectively \cite{nandori,nagys_plb}. The evolution of
the potential freezes out below the mass gap of the MSG model and a non-trivial
potential is left over in the IR end point, reflecting the state of affairs
at $k\approx M$. The IR scaling is trivial for
$k<M$, $\tilde u_n(k)\sim k^{-2}$. In order to go beyond the asymptotic
scaling analysis and to find out more precisely the changes brought by the
non-periodical mass term to the RG flow we have to rely on the numerical
solutions of the evolution equations.

\begin{figure}[th]
\includegraphics[width=8cm]{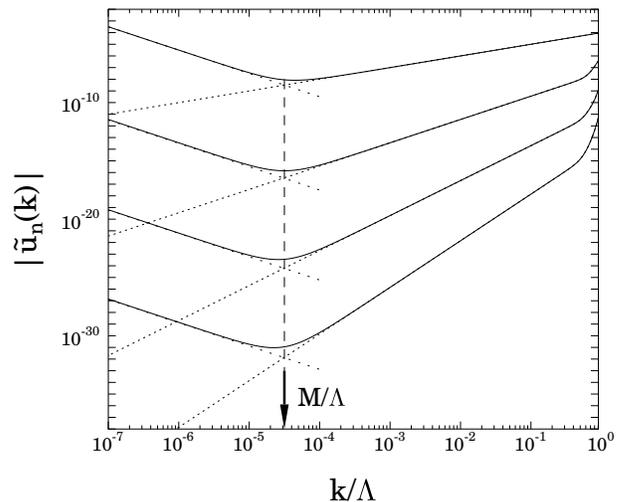}
\caption{RG flow of $\tilde u_1,\ldots,\tilde u_4$ for $\beta^2=12\pi$ and
$M^2=10^{-9}\Lambda^2$ (solid line) or $M^2=0$ (dense dotted line). The plot
also shows the asymptotic IR scaling $\sim k^{-2}$ (rare dotted line). 
The massive and the massless flows depart at the scale $M$,
indicated by a dashed vertical line, placed at the intersection of
the dotted lines.
\label{fig:sgmsg}}
\end{figure}

Let us consider first the regime $\beta^2>8\pi$ which is free of spinodal
instabilities in the massless case. The evolution of the first four
coupling constants, $\tilde u_1,\ldots,\tilde u_4$ is shown in \fig{fig:sgmsg}
for $\beta^2>8\pi$. The UV scaling regime is confined in this plot to the very 
beginning, around $k/\Lambda\approx1$ \cite{nagys_plb}, and what we see here is 
that the flows of the SG and the MSG models agree for $\beta^2>8\pi$ even in 
the IR scaling regime down to the mass gap. The freeze-out below the mass gap
takes place naturally without generating spinodal instabilities. 

\begin{figure}[t]
\includegraphics[width=8cm]{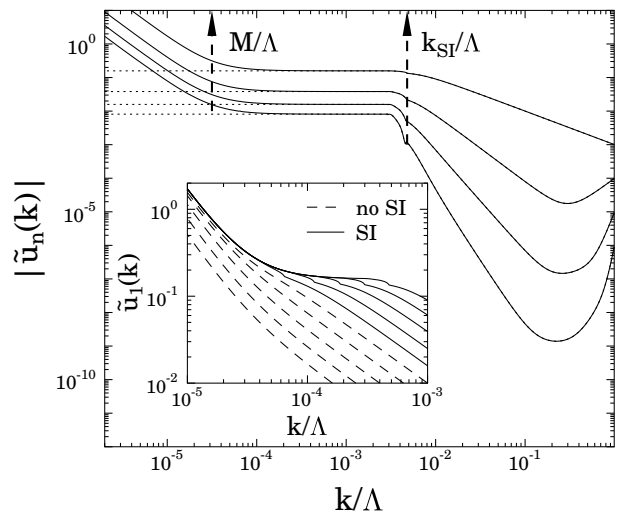}
\caption{RG flow of $\tilde u_1,\ldots,\tilde u_4$ for $\beta^2=4\pi$ and
$M^2=10^{-9}\Lambda^2$ (solid line) or $M^2=0$ (dotted line). The scales of
the spinodal instability $k_\mr{SI}/\Lambda$ and the mass $M/\Lambda$ are
indicated. The inset shows RG flows of $\tilde u_1$ for several different
initial values at $M^2=10^{-9}\Lambda^2$. 
\label{fig:treesgmsg}}
\end{figure}

The comparison of the massive and the massless cases is more involved for
$\beta^2<8\pi$ due to the appearance of instabilities. If the scale $k_\mr{SI}$
where instabilities appear is higher than the mass gap, $k_\mr{SI}>M$, then
the RG flows of the MSG and SG models are similar down to $M$ and they, i.e.
both display instabilities and differ for $k<M$ only, as shown in
\fig{fig:treesgmsg}. When the freeze-out scale is reached first during the
evolution, i.e. the scale $k_\mr{SI}$ of the SG model with the same potential
as $V(\phi)$ of the MSG model is smaller than $M$ then the instability does
not occur in the MSG model.

\begin{figure}[ht]
\includegraphics[width=8cm]{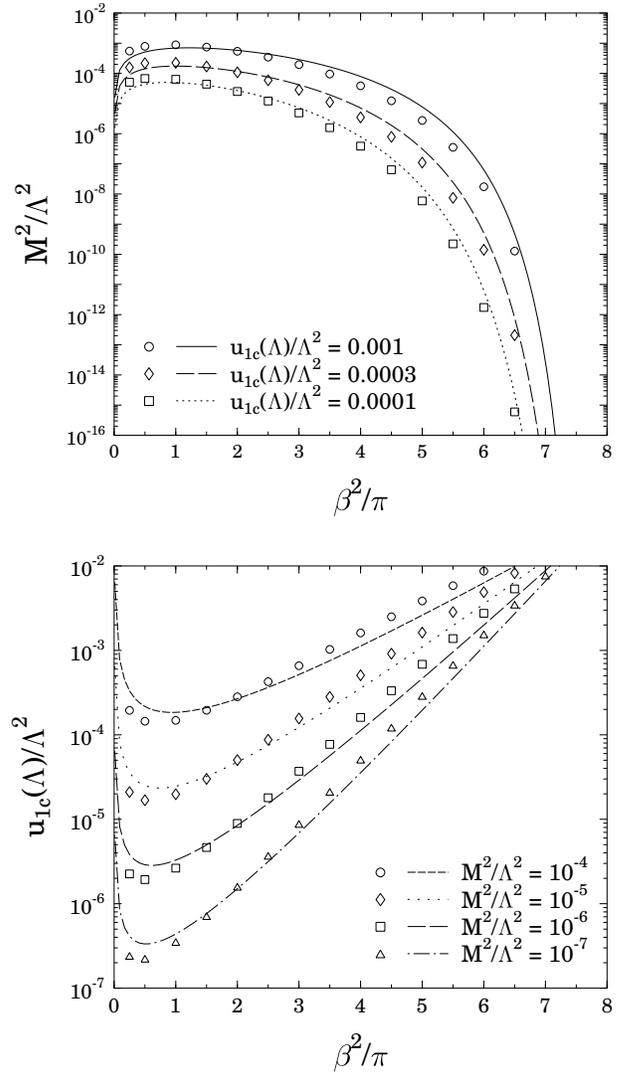}
\caption{The region with spinodal instability. Different symbols show the
boundary obtained by the numerical computation with 10 coupling constants and
the lines show the results obtained by taking into account a single Fourier
mode only.
\label{fig:phase}}
\end{figure}

It is instructive to determine the boundary of the region in the coupling
constant space with spinodal instability with the truncation where a single
Fourier mode is kept only, $\tu_n=\tu\delta_{1,n}$. The condensate appears
during the evolution at scale $k_\mr{SI}$, satisfying
$k^2_\mr{SI}+M^2+U''_{k_\mr{SI}}(\phi)=0$ for some $\phi$. The approximate
analytic expression, \eqn{nonlinsol1}, gives
\beq
k^2_\mr{SI} = (\Lambda^2+M^2)\left(\frac{\Lambda^2+M^2}{\beta^2 u(\Lambda)}
\right)^{\frac{8\pi}{\beta^2-8\pi}}-M^2
\label{ksi}
\eeq
for $\beta^2<8\pi$, suggesting that the coupling constant can be weak enough
to allow the mass term to remove the condensate. The RG flow obtained 
numerically for $\tilde u_1(k)$ shown in the inset of \fig{fig:treesgmsg}
confirms, as well, that the mass can be strong enough to prevent the formation
of instabilities. One can get an estimate of the critical value of the
coupling constant by equating $k_\mr{SI}$ and $M$ in \eqn{ksi},
\beq
u_{1c}(\Lambda) = \frac{\Lambda^2+M^2}{\beta^2}
\left(\frac{2M^2}{\Lambda^2+M^2}\right)^{1-\frac{\beta^2}{8\pi}}.
\label{u1c}
\eeq
The boundary of the region with instability is shown in \fig{fig:phase}.
In contrast to the SG model where the instability extends over the whole phase
with $\beta^2<8\pi$ the mass term always wins at the IR end point of the flow
of the MSG model and removes the condensate at some low but finite value of
the scale $k$.

The disappearance of spinodal instability and the trivial scaling,
$|{\tilde u}_n|\sim k^{-2}$ \cite{Ichinose} can be made plausible also by the
following, simple analytic consideration. Namely, \eqn{WHdim} can be rewritten
as
\beq
\sum_s {\cal B}_{n,s}k\partial_k \tu_s(k) =
\sum_s\left[-2{\cal B}_{n,s}+\frac{\alpha_2\beta^2 n^3 k^2}{k^2+M^2}\right]\tu_s(k),
\label{genv_kto0}
\eeq
with
\beq
{\cal B}_{n,s}=s\delta_{n,s}
-\frac{s\beta^2(n{-}s)^2\tu_{|n{-}s|}(0){-}(n{+}s)^2\tu_{n{+}s}(0)}{2(k^2+M^2)},
\eeq
resulting
\beq
{\cal B}_{n,s}\sim s\delta_{n,s}-\frac{s\beta^2a_{n,s}}{2M^2},
\eeq
an RG invariant quantity in the IR scaling region.
Finally, the second term on the right hand side of \eqn{genv_kto0} can be
neglected for $k\ll M$, yielding $\tu_n(k)\sim u_n(0)k^{-2}$ in the IR scaling
region.

\section{Phase structure}\label{sec:phase}

The mass term deforms the phase boundary of the SG model by extending the
ionized phase. In this phase of the SG model the IR scaling law generates the
scale dependence of the coupling constants $u_n(k)$ with $n\ge2$ through
$u_1(k)$, namely renders the ratios $R^{SG}_n=u_n(k)/u_1^n(k)$ RG invariant.
It was checked numerically that $R_n^{MSG}=|u_n(k)|/u_1^n(k)$ is RG invariant
in the IR scaling region of the MSG model without condensate and the potential
at $k=0$ depends on the initial value of $u_1$ only.

The phase with condensate is similar to those of the SG model. The potential
develops quickly into a super-universal, initial condition independent shape 
\cite{nagys_plb} when $M<k\approx k_\mr{SI}$, cf. the inset of
\fig{fig:treesgmsg}. But this scaling regime ends at $k\approx M$  where
trivial scaling laws come into force down to $k=0$. The matching of the
IR scaling of the SG model \cite{nagys_plb} with the trivial scaling law gives
$u_n(0)=(-1)^{n+1}2M^2/n^2\beta^2$.

\begin{figure}[ht]
\includegraphics[width=8cm]{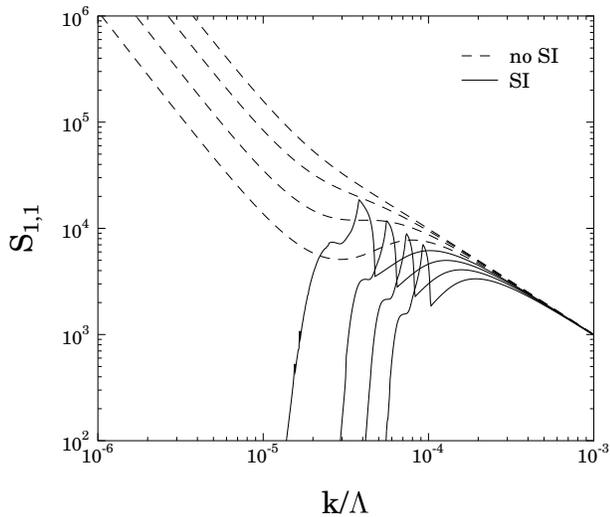}
\caption{The $(1,1)$ element sensitivity matrix is computed numerically 
for various initial conditions for $\beta^2{=}4\pi$ as the function of
$k/\Lambda$. The trajectories of molecular and the ionized phases are shown
by dashed and solid lines, respectively.
\label{fig:sens}}
\end{figure}

The modification of the phase boundary induced by the mass can be seen by means
of the sensitivity matrix \cite{janosRG,nagys_plb}, too. This matrix, defined as
the derivatives of the running coupling constants with respect to the bare one,
\beq\label{smatrix}
S_{n,m} = \frac{\partial \tu_n(k)}{\partial \tu_m(\Lambda)},
\eeq
develops singularities when the UV and IR cutoffs are removed at the phase 
boundaries only. The typical behaviour is depicted in \fig{fig:sens},
showing that the appearance of the condensate generates first singular
turns and leads later to radically different scale-dependence in this matrix.

The molecular phase of the two-dimensional SG model was compared with the
four-dimensional Yang-Mills theory \cite{nagys_plb} because both support
spinodal instabilities in their vacuum and the periodic symmetry of the SG
model is formally similar to the center symmetry of the gauge theory. This
analogy can be extended to the MSG model, where the breakdown of the
periodicity which manifests itself in the scaling laws for $k<M$ is comparable
with the spontaneous symmetry breaking in a four dimensional Yang-Mills-Higgs
system. The solitons cease to be stable and elementary plane-wave excitations
appear in the scattering matrix of the MSG model due to the breakdown of the
periodicity by the mass term. In a similar manner, the confining forces are
lost by the spontaneous breakdown of the global symmetry and elementary
excitations propagate in the vacuum of gauge theories. In other words, the
competition of the confinement and the symmetry breaking scales in the gauge
theory is similar to what happens between $k_\mr{SI}$ and $M$ in the MSG model.

\section{RG microscope}\label{sec:mic}

The non-triviality of the IR scaling regime opens the possibility of having
relevant  non-renormalizable operators, being irrelevant around the UV fixed
point, but becoming  relevant in the IR scaling regime. To find out whether
this can be the case let us take the running coupling constants, given by
\eqn{nonlinsol1}
\bea
\tu_n(k) &=&(-1)^{n+1}\tu_1^n(\Lambda) R^\mr{MSG}_n k^{2n-2}\nn
&&\times
\left(\frac{k}{\Lambda}\right)^{-2n}
\left(\frac{k^2 + M^2}
{\Lambda^2 + M^2}\right)^{\frac{n\beta^2}{8\pi}},
\eea
and calculate the $(1,1)$ element of the sensitivity matrix in \eqn{smatrix}
in the limits $k\to 0$ and $\Lambda\to\infty$,
\beq
S_{1,1}\sim k^{-2}\Lambda^{\frac{8\pi-\beta^2}{4\pi}}.
\eeq
We set $\beta^2>8\pi$, i.e. move deep in the ionized phase in the language
of the SG model where all coupling constants are non-renormalizable. But the
IR scaling law makes the coupling constants relevant with its factor $k^{-2}$.
More precisely, the dimension 2 coupling constants freeze out meaning that
their values counted in the `natural' units of the running cutoff, $k$, diverge
as the IR end point, $k=0$, is approached. The result is the regain of the
sensitivity of the long distance physics on the choice of the bare,
microscopic, non-renormalizable coupling constants which was suppressed in the
UV scaling regime. As long as the UV cutoff $\Lambda$ and the IR observational
scale $k$ are chosen according to $k=t^{-p_\mr{IR}}\mu$ and
$\Lambda=t^{p_\mr{UV}}\mu$ where $p_\mr{IR}=p_\mr{UV}(\beta^2/8\pi-1)>0$
(with any $t>1$ and $\mu>0$), the observed IR dynamics depends on the choice of
the bare coupling constant. This rearrangement is reminiscent of the
traditional microscope in the sense that the amplification, i.e. divergence of
the RG flow in the IR scaling regime balances the suppression, i.e. focusing
of the flow in the UV scaling regime. As a result the large scale  observations
can fix the value of the microscopic parameter with good accuracy. Notice that
this kind of dynamics must be put  in the theory at microscopic scale even
though it starts to influence the physics in the IR scaling regime. Such an
interplay between the different scaling regimes represents a highly
non-trivial, global extension of the simple universality idea which is based
on the local analysis of the RG flow at a given fixed point and
might be a key to phenomena like superconductivity and quark confinement
\cite{janosRG,grg}.

\section{Two phases of QED$_2$}\label{sec:qed}

The most remarkable consequence of the non-trivial phase structure
of the MSG model is that the
two-dimensional quantum electrodynamics (QED$_2$) also has two phases.
The Lagrangian of the QED$_2$ is given as
\beq
{\cal L}=-\frac14 F_{\mu\nu}F^{\mu\nu}+\bar\psi\gamma^\mu(\partial_\mu-
\ci e A_\mu)\psi-m\bar\psi\psi,
\label{mSch}
\eeq
where $F_{\mu\nu}=\partial_\mu A_\nu-\partial_\nu A_\mu$, $m$ and $e$ are the
bare rest mass of the electron and the bare coupling constant, respectively.
The bosonization rules gives
\bea
:\bar\psi\psi:&\to&-cmM\cos(2\sqrt{\pi}\phi),\nn
:\bar\psi\gamma_5\psi:&\to&-cmM\sin(2\sqrt{\pi}\phi),\nonu
j_\mu=:\bar\psi\gamma_\mu\psi:&\to&\frac1{\sqrt{\pi}}
\varepsilon_{\mu\nu}\partial^\nu\phi,\nn
:\bar\psi\ci\sla\partial\psi:&\to&\frac12 N_m (\partial_\mu\phi)^2,
\eea
where $N_m$ denotes normal ordering with respect to the fermion mass $m$,
$c=\exp{(\gamma)}/2\pi$ with the Euler constant $\gamma$, and $M=e/\sqrt{\pi}$ the `meson' mass.
The MSG Hamiltonian
\beq
{\cal H} = N_M\int_x\left[\frac12\Pi^2_x+\frac12(\partial_1\phi_x)^2+
\frac12M^2\phi_x^2+u_1\cos(\beta\phi_x)\right]
\label{hamscch}
\eeq
for $\beta^2=4\pi$ and $u_1=cmM$ is the bosonized version of the QED$_2$.
According to \eqn{dlmassev} the dimensionful electric charge $e = M\sqrt{\pi}$ does not evolve.
Using \eqn{nonlinsol1} the flow of the electron mass $m(k)$
\beq\label{emflow}
m(k) = m(\Lambda)\left(\frac{k^2 + M^2}{\Lambda^2 + M^2}\right)^{1/2}, \quad
m(\Lambda)\equiv\frac{u_1(\Lambda)\sqrt{\pi}}{c e},
\eeq
inherits all the properties of the flow of the
first Fourier amplitude $u_1(k)$. It implies that there exists a critical
value
\beq\label{me}
m_c(\Lambda) = \frac{u_c(\Lambda)\sqrt{\pi}}{c e}
=\frac{\sqrt{2}}{4\pi c}(\Lambda^2+e^2/\pi)^{1/2}
\eeq
of the bare mass which determines whether the IR value of the mass depends on its UV value or not.
For $m(\Lambda)>m_c(\Lambda)$ the evolution runs into the spinodal
instability and the IR value of $m$ becomes independent of its bare value:
\beq
m(k\to 0) = \frac{2e\sqrt{\pi}}{c 4\pi^2},
\eeq
while for $m(\Lambda)<m_c(\Lambda)$ \eqn{emflow} implies the IR behaviour
\beq\label{emflow2}
m(k\to 0) = m(\Lambda)\left(\frac{M^2}{\Lambda^2 + M^2}\right)^{1/2}.
\eeq
Therefore the sensitivity matrix in the parameter space $(m,e)$
taking the values $S_{1,1} = 0$ for $m(\Lambda)>m_c(\Lambda)$ and
\beq
S_{1,1} = \frac{\partial m(k)}{\partial m(\Lambda)} =
\left(\frac{k^2 + e^2/\pi}{\Lambda^2 + e^2/\pi}\right)^{1/2},
\eeq
for $m(\Lambda)<m_c(\Lambda)$ indicates the existence of two different
phases in QED$_2$.
Lattice calculations also affirmed this result
\cite{BySrBuHa2002,ShMu2005}.
For large coupling ($e >> m$), the model has a unique vacuum at 
$\varphi=0$. For weak coupling ($e << m$), the reflection
symmetry is spontaneously broken and the model has non-trivial
vacua, located approximately at $\phi = \pm \sqrt{\pi}/2$.
According to lattice simulations and density matrix RG studies
of the MSG model the critical value 
which separates the two phases of the model is $m/e_c=0.3335$.
The analytical result $m/e_c=0.3168$ for the critical point in
\eqn{me} suggests that the
RG methods using LPA enables us to  determine the
phase structure of the MSG model in a reliable manner.

\section{Internal space renormalization}\label{sec:internal}

One can go beyond the LPA by using the internal space RG method.
We define the generating functional of the connected Green functions as
\beq
W[j] = \log \int{\cal D}\phi e^{-S_B[\phi]+j\cdot\phi},
\eeq
with external source $j_x$. The shorthand notation $f\cdot g=\int_xf_xg_x$ is used.
The effective action is defined as the Legendre-transform of $W[j]$,
\beq
\Gamma[\varphi] = j\cdot\varphi-W[j],
\eeq
where the external source $j_x$ is expressed in terms of $\varphi_x$ according to the
 implicit equation
\beq
\varphi = \frac{\delta W[j]}{\delta j}.
\eeq
The idea of internal-space RG is to eliminate quantum fluctuations successively ordering them
according to their amplitudes. This can be achieved by introducing an additional mass term into the action,
\beq
S_\lambda[\phi] = S_B[\phi]+\hf \lambda^2 \phi^2
\eeq
with the control parameter $\lambda$. For $\lambda=\lambda_0$ being of the order of the UV 
cut-off $\Lambda$ the large-amplitude fluctuations are suppressed and decreasing 
the evolution parameter $\lambda$ towards zero, they are continuously accounted for.
Let us separate off the suppressing mass term from the evolving effective action,
\beq
\Gamma_\lambda[\varphi] ={\bar \Gamma}_\lambda[\varphi]
+\hf \lambda^2\varphi^2,
\eeq
and use the ansatz
\beq\label{eaans}
{\bar \Gamma}_\lambda[\varphi] 
= \int_x\left[\frac12 (\partial_\mu\varphi_x)^2+U_\lambda(\varphi_x)\right]
\eeq
with $U_\lambda(\varphi)=\hf M^2_\lambda\varphi^2+V_\lambda(\varphi)$ and
$V_\lambda(\varphi)=\sum_{n=1}^\infty u_n(\lambda)\cos(n\beta_\lambda\varphi)$.
The functional evolution equation is
\beq
\partial_{\lambda^2}{\bar \Gamma}_\lambda =
\hf\mr{Tr} \left[\lambda^2\delta_{x,y}+{\bar\Gamma}^{(2)}_{\lambda~x,y}\right]^{-1},
\eeq
where ${\bar\Gamma}^{(2)}_{\lambda~x,y}=
\delta^2{\bar\Gamma}_\lambda/\delta\varphi_x\delta\varphi_y$, \cite{internal}, reads
\beq\label{ea_f}
\partial_{\lambda^2} V_\lambda(\varphi) = \hf\int_{\bf p}\frac1{{\bf p}^2+\lambda^2+M^2_\lambda+V''_\lambda(\varphi)}
\eeq
for homogeneous field configurations $\varphi$
for the potential $V_\lambda(\varphi)$ of the ansatz  \eqn{eaans} where $V_\lambda''(\varphi)
=\partial^2V_\lambda(\varphi)/\partial\varphi^2$.
The two-dimensional momentum integral can easily be performed, giving
\beq\label{ea_func}
(1+\lambda^2\partial_{\lambda^2})\tilde V_\lambda
=\frac1{8\pi}\log\left[
\frac{(\Lambda/\lambda)^2+1+\tilde M^2_\lambda+\tilde V''_\lambda}
{1+\tilde M^2_\lambda+\tilde V''_\lambda}\right],
\eeq
where the dependences on the field variable $\varphi$ is suppressed.
By performing the Fourier expansion in both sides of \eqn{ea_func}  one obtains
a set of differential equations for the couplings $\tu_n(\lambda)$, $\beta_\lambda$
and $\tilde M_\lambda$. Since the left hand side of \eqn{ea_f} does not contain polynomial
terms, the mass parameter does not evolve, $M_\lambda^2=M^2$.
Thus the mass is a relevant parameter of the LPA ansatz for all scales,
\beq
\tilde M^2_\lambda=\lambda^{-2} M^2,
\label{dlmassev_eff}
\eeq
cf. \eqn{dlmassev}. Furthermore, as in the case of the WH-RG equations,
if the right hand side at a given scale $\lambda$ has a given period $2\pi/\beta_\lambda$
in the internal space, then $V_{\lambda-d\lambda}$ shall have the same period, so that
the parameter $\beta_\lambda=\beta$ evolves neither. 
Note that the argument of the logarithm on the r.h.s.
of \eqn{ea_func} should be positive, otherwise the  evolution
fails. This criterion is reminiscent of the WH-RG method and it
may generate spinodal instability like singularities on the internal-space RG renormalized trajectory,
to be interpreted as quantum phase transition.
In order to obtain the evolution equations for the couplings more easily
we differentiate \eqn{ea_func} with respect to $\varphi$,
\bea
&&-\frac1{8\pi}\frac{\Lambda^2}{\lambda^2}\tilde V_\lambda''' =\nn
&&~\left(\frac{\Lambda^2}{\lambda^2}+1+\tilde M^2+\tilde V_\lambda''\right)(1+\tilde M^2+\tilde V_\lambda'')
(1+\lambda^2\partial_{\lambda^2})\tilde V_\lambda'.\nn
\eea
The evolution should be started at $\lambda_0^2\sim \ord{\Lambda^2}\gg 
M^2\gg |V_\lambda''|$. At these 'UV' scales the large-amplitude field fluctuations with $|\varphi|^2\gg
|V_\lambda''|/ \lambda^2$ are suppressed and the effective action can be calculated perturbatively.  

\subsection{Asymptotic scaling}

The asymptotic scaling for $\lambda^2$ satisfying $\Lambda^2\gg \lambda^2\gg 
M^2\gg |V_\lambda''|$ can be deduced by using the independent mode approximation,
\beq
(1+\lambda^2\partial_{\lambda^2})\tilde V_\lambda =
-\frac1{8\pi}\left[\frac1{1+\tilde M^2}-\frac1{\Lambda^2/\lambda^2+1+\tilde M^2}\right]\tilde V'',
\eeq
where the field independent terms are neglected.
The evolution of the couping constants decouple and one finds
\beq\label{tu_1_eff}
\tu_n(\lambda) = \tu_n(\lambda_0)\left(\frac{\lambda}{\Lambda}\right)^{-2}
\left(\frac{\lambda^2+M^2}{\Lambda^2+\lambda^2+M^2}\right)^
\frac{n^2\beta^2}{8\pi}
\eeq
for the dimensionless couplings constants $\tu_n(\lambda)$, yielding
\beq\label{asysc}
\tu_n\sim \lambda^{n^2\beta^2/4\pi-2}.
\eeq

\subsection{Ionized phase}

In order to find out the non-asymptotical regime one has to solve a system of
coupled evolution equations for the couplings constants $\tu_n(\lambda)$ as the control 
parameter $\lambda$ is decreased from $\lambda_0=\Lambda$ down to $\lambda=0$. 
The evolution equations has been derived  and solved for $\beta^2>8\pi$ numerically.
It was found that the increase of the number of the  couplings $ \tu_n$ beyond $n=10$
does not influence the evolution of the first few couplings, similarly to the
WH-RG equations.

The evolution of the first four
coupling constants, $\tilde u_1,\ldots,\tilde u_4$ is shown in \fig{fig:sgmsgeff}
for $\beta^2=12\pi$. The internal space RG method gives qualitatively the same
scaling laws both in the 'UV' and in the 'IR' regions just as the WH-RG method.
The numerical value of the fundamental coupling constant $\tu_1(\lambda)$
was found to follow closely the analytic form of \eqn{tu_1_eff}. The renormalized
trajectory shares the feature, known from the WH-RG scheme that
the SG and the MSG models with $\beta^2>8\pi$ agree in the IR scaling regime down to the mass gap.

\subsubsection{IR scaling}

\begin{figure}[th]
\includegraphics[width=8cm]{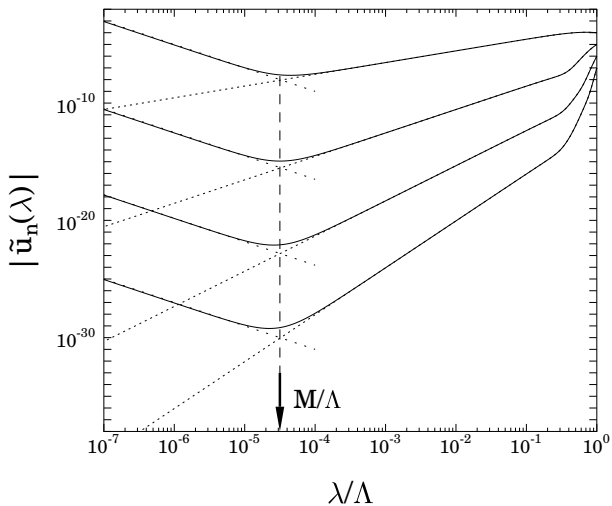}
\caption{RG flow of $\tilde u_1,\ldots,\tilde u_4$ for $\beta^2=12\pi$ and
$M^2=10^{-9}\Lambda^2$ (solid line) or $M^2=0$ (dense dotted line). The plot
also shows the asymptotic IR scaling $\sim \lambda^{-2}$ (rare dotted line). 
The massive and the massless flows depart at the scale $M$,
indicated by a dashed vertical line, placed at the intersection of
the dotted lines.
\label{fig:sgmsgeff}}
\end{figure}

The decrease  of the control parameter $\lambda$ drives us out from
the asymptotic region. According to
\fig{fig:sgmsgeff} the IR scaling is
$\tu_n\sim\lambda^{n(\beta^2/4\pi-2)}$. Such a scaling behaviour
can be obtained analytically from the functional RG equation
in \eqn{ea_func} which becomes
\beq
(1+\lambda^2\partial_{\lambda^2})\tilde V_\lambda'
+\tilde V_\lambda''(1+\lambda^2\partial_{\lambda^2})\tilde V_\lambda'
= -\frac1{8\pi}\tilde V_\lambda'''
\eeq
for $M^2\ll\lambda^2\ll\Lambda^2$. Making the ansatz
\beq\label{tu_n_ans}
\tu_n(\lambda) = c_n \lambda^{n\eta}
\eeq
with $\eta\ge 0$. For $n=1$ one gets $\eta =\beta^2/4\pi-2>0$. For
$n>1$ we obtain the recursion relation
\beq
c_n=\frac{\hf\beta^2 \sum_{s=1}^{n-1}  (2+s\eta)s(n-s)^2c_{n-s}c_s}
{n(2+n\eta-n^2\frac{\beta^2}{4\pi})}.
\label{c_n}
\eeq
The coefficients $c_n$ can be expressed in terms of $c_1$, since
$c_1=\tilde u_1(\Lambda)(\lambda/\Lambda)^\eta$ and therefore
$c_n=(-1)^{n+1}\tilde u_1^n(\Lambda)R_n$, with $R_1=1$ and all  $R_n$
being independent of the bare couplings. These properties were confirmed numerically.
They imply that the dimensionless effective action in the IR regime can be
parametrized by the single bare parameter $\tu_1(\lambda)$.

In order to consider the effect of the mass $M$ in the
theory we also determined the RG evolution of the couplings  for the (massless) SG model,
which depicted  by dense dotted lines in \fig{fig:sgmsgeff}.
One finds that the  scaling behaviour in \eqn{tu_n_ans} goes on to
infinitesimal values of $\lambda$. We note that all couplings are irrelevant
and the effective action is zero in the $\lambda\to 0$ limit.
After introducing the mass $M$ it is clear from \fig{fig:sgmsgeff}
that 
the evolution of the potential freezes out below the mass gap and a non-trivial dimensionless
potential is left over in the IR end point. The IR scaling is trivial for
$\lambda<M$, $\tu_n(\lambda)\sim \lambda^{-2}$.
Since the evolutions for the SG and MSG models are identical down to the scale
$\lambda\sim M$,
the evolving effective action of the MSG model inherits the properties of the  SG model, namely
it depends on the initial value of the fundamental mode $\tu_1(\lambda_0)$ only.
In fact, it was found 
numerically that $R_n^{MSG}=|u_n(\lambda)|/u_1^n(\lambda)$ is RG invariant
in the scaling region $\lambda< M$.

One sees also that for $\beta^2>8\pi$ theories with various values of the mass $M$
belong to the same phase. This is conclusion arises because one detects the same scaling behaviour of the
dimensionless parameters $\tilde u_n(\lambda)$ for
$\lambda/M<1$ and the same qualitative behaviour of the RG invariant constants
$R_n^{MSG}$ for all values of $M$. 

\subsection{Molecular phase}

\begin{figure}[t]
\includegraphics[width=8cm]{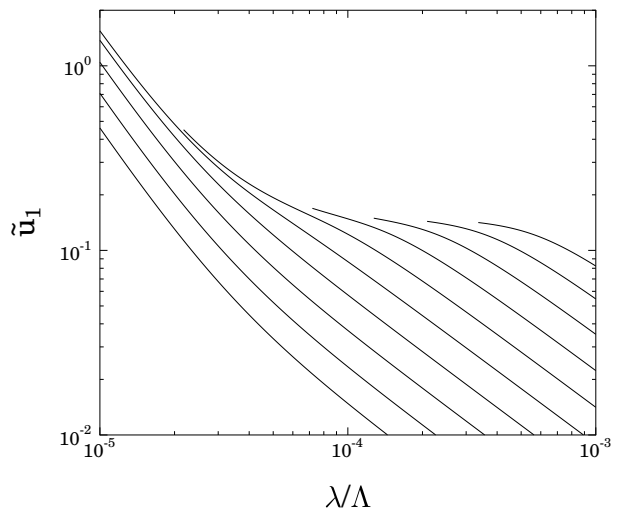}
\caption{Internal-space RG evolution of the coupling $\tilde u_1$ for several different
initial values  at $M^2=10^{-9}\Lambda^2$.
\label{fig:u1sens}}
\end{figure}

The typical evolution is
depicted for several bare values  $\tu_1(\Lambda)$ in \fig{fig:u1sens} for $\beta^2<8\pi$.
Numerics shows immediately, that the evolution stops at a non-vanishing value of $\lambda=\lambda_c$ due to
the appearing a  negative argument of the logarithm in
\eqn{ea_func} for the  $\tu_1(\Lambda)> \tu_{1c}$, where $\tu_{1c}$ is a critical value of the
first Fourier amplitude of the bare periodic potential. Then the evolution equation
loses its validity and presumably
an alternative RG equation is necessary as in  the WH-RG framework,
where  the appearance of the spinodal instability implies a tree-level
blocking relation \cite{nandori,janosRG,nagys_plb}. 
Though plausible but is not obvious that the  singularity in the internal-space evolution
is also rooted in the spinodal instability.
The clarification of this point needs further efforts.

For bare values $\tu_1(\Lambda)<\tu_{1c}$
the evolution does not stop and goes below  the mass scale.  Then, as in the
case of $\beta^2>8\pi$, the trivial scaling $\tu_n(\lambda)\sim \lambda^{-2}$
is obtained in the limit $\lambda \to 0$.

\begin{figure}[ht]
\includegraphics[width=4.5cm,angle=-90]{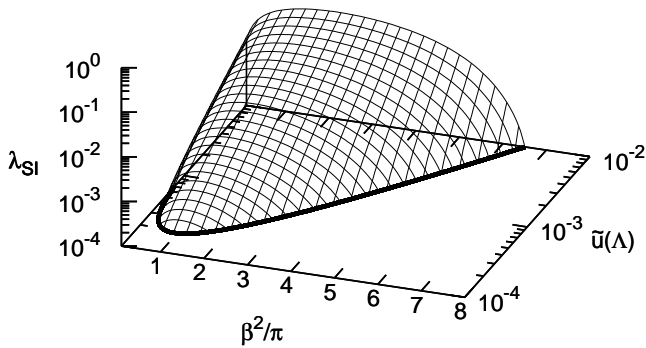}
\caption{The scale $\lambda_c$, where the singularity appears, as the function of the
initial value of the first Fourier amplitude $u_{1c}$ and $\beta^2/\pi$ for $M^2=10^{-4}$.
\label{fig:phaseeff}}
\end{figure}

Although one cannot follow the RG evolution for $\lambda<\lambda_c$ for $\tu_1(\Lambda)>\tu_{1c}$
by means of \eqn{ea_func},
one can still determine the critical value $\tu_{1c}$ analytically, when   the local periodic 
potential is restricted to its first Fourier mode. The singularity appears
during the evolution at the scale $\lambda_c$ satisfying
$\lambda^2_C+M^2+V''_{\lambda_c}(\varphi)=0$ for some $\varphi$.
 Using \eqn{tu_1_eff} as a good approximation of the scaling for $\lambda>\lambda_c$,
one finds
\beq
\lambda^2_c = \Lambda^2\left(\beta^2 \tu_1(\Lambda)
\right)^{\frac{\beta^2-8\pi}{8\pi}}-M^2
\label{ksiL}
\eeq
for $\beta^2<8\pi$. The  negativity of the right hand side suggests that the coupling
 constant is sufficiently  weak 
to allow the mass term to remove the singularity.
For the opposite case $\lambda_c^2>0$ one can  estimate  the critical value of the
coupling constant by equating $\lambda_c$ to $M$ in \eqn{ksiL},
\beq
\tilde u_{1c} = \frac1{\beta^2}
\left(\frac{2M^2}{\Lambda^2}\right)^{1-\frac{\beta^2}{8\pi}}.
\label{u1c2}
\eeq
The value of $\lambda_c$ is plotted on the plane $(\beta^2/\pi,\tu_1(\Lambda))$
in Fig. \ref{fig:phaseeff}. In contrast to the SG model where the singularity appears
for all $\tu_1(\Lambda)$ for  $\beta^2<8\pi$ the mass term always wins at the IR end
 point of the evolution
for the MSG model and removes the singularity  at some low but finite 
scale $\lambda_c$.

The introduction of the wave function renormalization might modify the
phase structure. Nevertheless, one expects that both `molecular'  phases
survive wave-function renormalization. This is  because the MSG model
with the particular value of $\beta^2=4\pi$
represents  the bosonized version of the 2-dimensional
quantum electrodynamics QED$_2$ which exhibits two phases.

\section{Summary}\label{sec:sum}

Some global features of the RG flow of the MSG model are discussed in this work.
It is shown that the model possesses condensate of elementary excitations
with non-vanishing momentum, spinodal instability, for weak enough mass
in the remnant of the molecular phase of the SG model. This condensate,
the sign of the periodicity of the local potential, generates non-trivial 
effective potential and phase structure despite the explicit, stable mass 
for elementary excitations in the deep IR region. The sensitivity matrix
allows us to study the way the ultraviolet parameters influence the IR physics.
It was found that the suppression of the sensitivity on the nonrenormalizable
bare coupling constants, generated in the UV scaling regime, can be overturned
by the increasing sensitivity piled up in the IR scaling regime if the UV and
the IR cutoffs are removed in a coordinated manner. As a result, a non-trivial,
global extension of the universality is found which goes beyond the local
studies of the RG flow around the UV fixed point only.

\section*{Acknowledgement}

This work has been supported by the Grant MTA-CNRS. S.N.
acknowledges the Grant \"Oveges of the National Office for Research and Technology,
and the Grant of Universitas Foundation, Debrecen.


\begin{thebibliography}{99}
\bibitem{univ} K.\ G.\ Wilson, J. Kogut, Phys. Rep. C{\bf 12}, 77 (1974);
K.\ G.\ Wilson, Rev. Mod. Phys. {\bf  47}, 773 (1975);
Rev. Mod. Phys. {\bf 55}, 583 (1983).
\bibitem{grg} J.\ Alexandre, V. Branchina, J. Polonyi, Phys. Rev. D {\bf 58}, 16002 (1998).
\bibitem{janosRG} J.\ Polonyi, Central Eur.J.Phys. {\bf 1}, 1 (2004).
\bibitem{Coleman_sg} S.\ R.\ Coleman Phys. Rev. D{\bf 11}, 2088 (1975);
AI. B. Zamolodchikov, Int. J. Mod. Phys. A {\bf 10}, 1125 (1995);
J. Balogh, A. Heged\H{u}s, J. Phys. A {\bf 33}, 6543 (2000); 
G. v. Gersdorf, C. Wetterich, Phys. Rev. B {\bf 64}, 054513 (2001); 
M. Faber, A. N. Ivanov, J. Phys. A {\bf 36}, 7839 (2003);
H. Bozkaya, M. Faber, A. N. Ivanov, M. Pitschmann, 
J. Phys. A {\bf 39}, 2177 (2006).
\bibitem{huang} K.\ Huang, J. Polonyi, Int. J. of Mod. Phys. {\bf 6},
409 (1991).
\bibitem{Coleman} S.\ R.\ Coleman, Annals Phys. {\bf 101}, 239 (1976);
S.\ R.\ Coleman {\it et al.}, Annals Phys. {\bf 93}, 267 (1975);
W.\ Fischler {\it et al.}, Phys. Rev. D{\bf 19}, 1188 (1979);
H. J. Rothe, K. D. Rothe, J. A. Swieca, Phys. Rev. D{\bf 19} (1979) 3020;
S.\ Nagy, J. Polonyi, K. Sailer, Phys. Rev. D{\bf 70}, 105023 (2004).
\bibitem{Ichinose} I.\ Ichinose, H. Mukaida, Int. J. Mod. Phys. A{\bf 9}, 1043
(1994); S.\ W. Pierson, O. T. Walls, Phys. Rev B {\bf 61}, 663 (2000).
\bibitem{nagys_jpa} S.\ Nagy, J. Polonyi, K. Sailer, J. Phys. A{\bf 39}, 8105 (2006).
\bibitem{nandori_npb} I.\ N\'andori, S. Nagy, K. Sailer, U. D. Jentschura,
Nucl. Phys. B {\bf 725}, 467 (2005).
\bibitem{Lu2000}
Wen-Fa Lu, Phys. Rev. D {\bf 59}, 105021 (1999);
J. Phys. G {\bf 26}, 1187 (2000).
\bibitem{wh}F.\ J.\ Wegner, A.\ Houghton, Phys. Rev. A.{\bf 8}, 401 (1973).
\bibitem{effac_rg}
J. Polchinski, Nucl. Phys B {\bf 231}, 269 (1984);
C. Wetterich, Phys. Lett. B {\bf 301}, 90 (1993);
N. Tetradis, C. Wetterich, Nucl. Phys B{\bf 398}, 659 (1993);
T. R. Morris, Int. J. Mod. Phys. A {\bf 9}, 2411 (1994);
D. F. Litim, Phys. Lett. B {\bf 486}, 92 (2000); 
J. Berges, N. Tetradis, C. Wetterich, Phys. Rept. {\bf 363}, 223 (2002);
H. Gies,  e-print: hep-ph/0611146;
D. F. Litim, JHEP 0507 (2005) 005;
D. F. Litim, J. M. Pawlowski, L. Vergara, e-print: hep-th/0602140.
\bibitem{internal} J.\ Alexandre, J.\ Polonyi, Annals Phys. {\bf 288}, 37 (2001);
J.\ Alexandre, J.\ Polonyi, K. Sailer, Phys. Lett. B{\bf 531}, 316 (2002).
\bibitem{nagys_plb} S. Nagy, I. N\'andori, J. Polonyi, K. Sailer,
Phys. Lett. B{\bf 647}, 152 (2007).
\bibitem{nandori} I.\ N\'andori, J. Polonyi, K. Sailer, Phys. Rev. D{\bf 63},
045022 (2001).
\bibitem{janostree} J.\ Alexandre, V.\ Branchina, J.\ Polonyi,
Phys. Lett. B{\bf 445}, 351 (1999); J.\ Polonyi, e-print: hep-th/0509078.
\bibitem{BySrBuHa2002}
T. M. Byrnes, P. Sriganesh, R. J. Bursill and C. J. Hammer,
Nucl. Phys. B (Proc. Suppl.) {\bf 109A}, 202 (2002);
Phys. Rev. D {\bf 66}, 013002 (2002).
\bibitem{ShMu2005}
R. Shankar, G Murthy, e-print: cond-mat/0508242.
\end{thebibliography}
\end{document}